\documentclass[9pt,twocolumn,twoside,superscriptaddress,prl]{revtex4-1}

\usepackage[utf8]{inputenc}
\usepackage[T1]{fontenc}

\usepackage[usenames,dvipsnames]{xcolor}

\newcommand{\orange}[1]{\textcolor{black}{#1}}
\newcommand{\blue}[1]{\textcolor{black}{#1}}

\usepackage[pdftex, hidelinks]{hyperref}

\newcommand\myshade{85}
\colorlet{mylinkcolor}{violet}
\colorlet{mycitecolor}{YellowOrange}
\colorlet{myurlcolor}{Aquamarine}
\hypersetup{
  linkcolor  = mylinkcolor!\myshade!black,
  citecolor  = mycitecolor!\myshade!black,
  urlcolor   = myurlcolor!\myshade!black,
  colorlinks = true,
}

\usepackage{amsmath}

\usepackage{amssymb}
\usepackage[separate-uncertainty=true,multi-part-units=single]{siunitx}
\usepackage[normalem]{ulem}

\newcommand{\mref}[2]{\hyperref[#1]{\ref*{#1}(#2)}}

\usepackage{graphicx}
\graphicspath{./img/}

\usepackage{dcolumn}
\usepackage{bm}
\usepackage{soul}

\newcommand{\Gopt}{\Gamma_\mathrm{opt}}
\newcommand{\GS}{\Gamma_\mathrm{S}}
\newcommand{\GAS}{\Gamma_\mathrm{AS}}
\newcommand{\Gm}{\Gamma_\mathrm{m}}
\newcommand{\Om}{\Omega_\mathrm{m}}
\newcommand{\nth}{\bar{n}_\mathrm{th}}
\newcommand{\ncav}{\bar{n}_\mathrm{cav}}
\newcommand{\nest}{\bar{n}_\mathrm{est}}
\newcommand{\Cq}{C_\mathrm{q}}
\newcommand{\nba}{\bar{n}_\mathrm{ba}}
\newcommand{\kf}{\kappa_\mathrm{f}}
\newcommand{\gtwo}{g^{(2)}}
\newcommand{\Df}{\Delta_\mathrm{f}}
\newcommand{\kB}{k_\mathrm{B}}

\newcommand{\aan}{\hat{a}}
\newcommand{\acr}{\hat{a}^{\dagger}}
\newcommand{\ban}{\hat{b}}
\newcommand{\bcr}{\hat{b}^{\dagger}}

\begin{document}

\title{Phonon counting thermometry of an ultracoherent membrane resonator near its motional ground state}

\author{I. Galinskiy}
\email{ivan.galinskiy@nbi.ku.dk}
\affiliation{Niels Bohr Institute, University of Copenhagen, Blegdamsvej 17, 2100 Copenhagen, Denmark}

\author{Y. Tsaturyan}
\altaffiliation{Present address: Pritzker School of Molecular Engineering, University of Chicago, Chicago, IL 60637, USA}
\affiliation{Niels Bohr Institute, University of Copenhagen, Blegdamsvej 17, 2100 Copenhagen, Denmark}

\author{M. Parniak}
\affiliation{Niels Bohr Institute, University of Copenhagen, Blegdamsvej 17, 2100 Copenhagen, Denmark}

\author{E.~S. Polzik}
\affiliation{Niels Bohr Institute, University of Copenhagen, Blegdamsvej 17, 2100 Copenhagen, Denmark}

\begin{abstract}
Generation of non-Gaussian quantum states of macroscopic mechanical objects is key to a number of challenges in quantum information science, ranging from fundamental tests of decoherence to quantum communication and sensing. Heralded generation of single-phonon states of mechanical motion is an attractive way towards this goal, as it is, in principle, not limited by the object size. Here we demonstrate a technique which allows for generation and detection of a quantum state of motion by phonon counting measurements near the ground state of a \SI{1.5}{\MHz} micromechanical oscillator. We detect scattered photons from a membrane-in-the-middle optomechanical system using an ultra-narrowband optical filter, and perform Raman-ratio thermometry and second-order intensity interferometry near the motional ground state ($\bar{n}=0.23\pm0.02$~phonons). \orange{With} \blue{an} \orange{effective mass in the} \blue{nanogram} \orange{range, our system lends itself for} \blue{studies} \orange{of long-lived non-Gaussian motional states} \blue{with} \orange{some of the heaviest objects to date.} 
\end{abstract}

\maketitle

\section{Introduction}
Single photon Raman scattering from a system initiated at or near its quantum ground state is a powerful method for generation of highly non-classical states. Addition or subtraction of a countable number of excitations is a common way to generate Fock states or Schrödinger cat states \cite{McConnell2015,Duan2001,Gerrits2010,Cooper2013,Andersen2013}. Of particular interest are situations where such operations are carried out by optical photon scattering from material systems with long coherence times, combined with high photon counting efficiency \cite{Bimbard2014}, since the memory capability effectively converts the heralded system into a deterministic source of single quanta.
Indeed, heralded schemes have been used to great effect in atomic systems \cite{Kuzmich2003,vanderWal196}, where fidelities and generation rates have improved steadily over the past few decades \cite{Zhao2009,Farrera2016,Parniak2017,Zugenmaier2018}. In the meantime, quantum systems based on mechanical resonators have shown great promise in the context of quantum transduction and communication, with devices exhibiting millisecond coherence times emerging in recent years \cite{Rossi2018}. As a natural step in expanding the toolbox available to cavity optomechanical systems, heralded schemes have recently been developed and implemented for mechanical devices \cite{Riedinger2016,Hong2017}, serving as a fundamentally new source of single quanta, namely phonons. \orange{In parallel, recent advances in quantum electromechanics have also allowed for generation of highly non-classical states of motion \cite{Clerk2020}.} Apart from applications in quantum information processing, where phonons can be mapped onto flying qubits, such non-Gaussian states of macroscopic mechanical resonators \cite{Khalili2010} have been suggested as a way to study gravitational decoherence processes \cite{Pepper2012,Weaver2018,Sekatski2014,PhysRevLett.113.020405}.

Here we report on a system which combines a number of versatile capabilities relevant for those applications. We realize phonon counting measurements \cite{Cohen2015} of a single high-$Q$ mechanical mode of motion of a membrane resonator \cite{Thompson2008,Tsaturyan2017,Nielsen2017,Rossi2018} \orange{in an optical cavity (see Fig. \ref{fig:expset})}. 
Thanks to cavity-based engineering of the optomechanical coupling, the interaction of light and mechanics is effectively dominated by a beamsplitter-like interaction between phonons and anti-Stokes photons:
\begin{equation}
    \hat{H}_\mathrm{BS} \propto \ban \acr_\text{AS} + \bcr \aan_\text{AS} \:,
\end{equation}
which leads to an exchange of optical and mechanical quanta, equivalent to anti-Stokes scattering of pump photons. This strong conversion of phonons into photons is the mechanism that both cools the mechanical resonator and maps its state onto light. Due to finite sideband resolution of the optomechanical cavity, there exists a small amount of two-mode squeezing interaction between phonons and Stokes photons ($\propto \bcr \acr_\text{S} + \ban \aan_\text{S}$),  as illustrated in Fig. \mref{fig:expset}{d}. This process introduces a small amount of heating that limits the minimum possible occupation of the mechanical oscillator under optical cooling \cite{Peterson2016}. In practice, however, frequency-resolved detection of Stokes and anti-Stokes photons, particularly for low-frequency mechanical oscillators, is challenging due to the presence of numerous nearby mechanical modes and a strong optical pump field.

Here we address this challenge with an ultra-narrowband spectral filter based on four cascaded free-space Fabry-Pérot cavities. This filter provides extremely efficient suppression ($>\SI{155}{\dB}$) of the Raman pumping light combined with highly selective detection of Stokes or anti-Stokes photons (Fig. \mref{fig:expset}{a,b}) detuned from the pump laser by the mechanical frequency. Importantly, the filter system efficiently suppresses other sources of spurious signals, such as nearby mechanical modes.

The excellent passive stability and high transmission of the system (30\% through the cascaded filter system), as well as robust and easily reproducible implementation of the filter cavity design, allow us to select sideband photons scattered by a single high-$Q$ mechanical mode. In parallel, we are able to perform high-efficiency heterodyne detection of sideband photons, thus demonstrating capabilities for versatile state engineering and characterization in both continuous-variable and discrete-variable domains.
Using the narrowband properties of the filter system, we perform Raman-ratio thermometry of a membrane resonator \cite{Purdy2015} by counting scattered Stokes- and anti-Stokes photons, and demonstrate the efficiency of this method for characterizing the effective mode temperature and coherence properties near the motional ground state. Finally, we analyze statistical properties of the Raman-scattered light and show single-mode thermal statistics with a coherence time matching the dynamical optical broadening of the mode, thus verifying its spectral purity.

The mechanical system employed here is a soft-clamped membrane resonator \cite{Tsaturyan2017} which has recently emerged as a viable platform for generation and storage of long-lived quantum states \cite{Rossi2018}. Already at moderate cryogenic temperatures, these resonators show millisecond coherence times. When combined with high detection efficiencies \cite{Nielsen2017}, which is necessary for optical quantum state tomography \cite{RevModPhys.81.299}, a membrane-in-the-middle system based on soft-clamped resonators lends itself ideal for studies of non-Gaussian states of motion \cite{Galland2014}. However, the inherent multimode nature of these resonators poses a major challenge in this pursuit, since mechanical modes in close proximity to the mode of interest (less than \SI{50}{\kHz} away) require strong spectral filtering.
As a consequence, all major experiments in the \si{\MHz}-frequency regime have thus far relied on homodyne or heterodyne detection, and photon counting techniques in optomechanics remained exclusive to the \si{\GHz}-frequency systems~\cite{Cohen2015}.

\section{Experimental setup}
\subsection{Optomechanical system}
\label{sec:omset}
\begin{figure}[ht]
\centering
	\includegraphics[width=\linewidth]{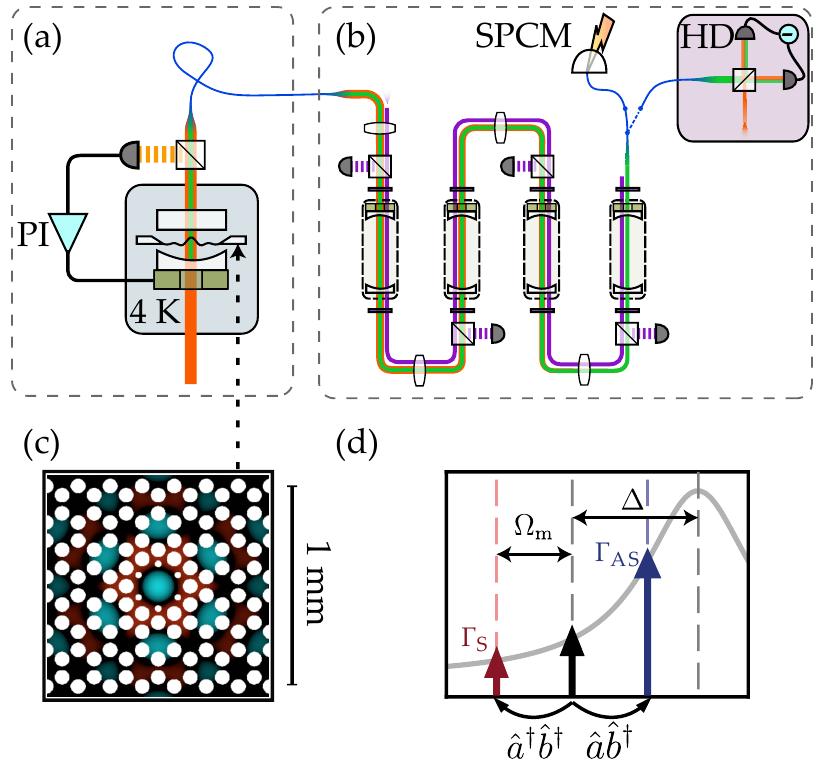}
	\caption{Experimental setup showing (a) the optomechanical cavity optically coupled to (b) four narrowband filtering cavities and subsequently directed to the single-photon detector (SPCM) or a balanced heterodyne detector (HD).  The optomechanical cavity is locked using a small portion of transmitted drive light (orange), with feedback provided by a PI controller connected to a piezo transducer. Filtering cavities are locked using an auxiliary beam (violet), as described in Supplementary Material. The filtered light (green) can be detected with photon counting (SPCM) or heterodyne detection (HD). Panel (c) shows the structure of the membrane with the defect in the center of the phononic crystal structure containing the high-$Q$ mode of interest. Colormap represents the displacement amplitude of the mode, with green and red corresponding to opposite signs of the displacement. In (d) we show the relevant frequencies with respect to the optical resonance (gray): optical drive (black) and Stokes and anti-Stokes sidebands (red and blue, exaggerated in comparison with drive).}
	\label{fig:expset}
\end{figure}

In the following section we describe our implementation of a system enabling phonon counting in the MHz-frequency regime.
Here, we use a soft-clamped silicon nitride membrane resonator as our mechanical system. The membrane resonator is \SI{12}{\nm}-thick, square-shaped with \SI{3.1}{\mm}-long edges, and is patterned with a phononic crystal structure, which includes a defect  ($\sim\SI{200}{\um}$ in diameter) in the center of the membrane \cite{Tsaturyan2017}. As shown in Fig. \mref{fig:expset}{c}, the defect hosts a localized radial vibrational mode at a mechanical frequency of $\Om/2\pi = \SI{1.48}{\MHz}$, which lies within the bandgap of the phononic structure. Soft-clamping and dissipation dilution provide a mechanical quality factor $Q = \Om/\Gamma_\mathrm{m}=\num{380(10)d6}$ for this defect mode, which has an effective mass of $\sim \SI{2}{\ng}$. The membrane is inserted into a high-finesse ($\mathcal{F}\approx22000$, linewidth $\kappa/2\pi=\SI{2.75}{\MHz}$) optical cavity with single-photon coupling rate of $g_0/2\pi \approx \SI{50}{\Hz}$. \orange{The cavity is placed} \blue{inside} \orange{a liquid helium flow cryostat. Under these conditions, }the linewidth of the cavity is sufficiently small to allow for ground-state \orange{sideband} cooling, where a minimum achievable occupation of $\sim 0.18$ phonons is given by the back-action limit \cite{Aspelmeyer2014}.

The cavity is pumped at a wavelength of $\lambda \sim \SI{852}{\nm}$ using a continuous-wave Ti:Sapphire laser (M Squared SolsTiS) allowing low-phase-noise operation. The laser drive is detuned to the red side of the cavity response in order to enable cooling and readout \cite{Aspelmeyer2014}.

Our optomechanical membrane-in-the-middle system allows reaching quantum cooperativities of $\Cq=4g^2/\Gamma_\text{m} \kappa \nth \sim100$ \cite{Rossi2018}, with $\nth$ being the phonon bath occupation and $g$ being the light-enhanced optomechanical coupling $g = g_0 \sqrt{\ncav}$, where $\ncav$ is the intracavity photon number of the drive light. In our case the bath phonon number is $\nth\approx\kB T/\hbar \Om\approx \num{1.3d5}$ due to the membrane substrate thermalization temperature estimated to be $T\approx\SI{9}{\kelvin}$. \orange{Combined with} \blue{a long energy decay time} \orange{of $T_1=1/\Gamma_\mathrm{M}\approx\SI{40}{\s}$, we}  \blue{estimate a decoherence} \orange{time of $T_2=1/(\bar{n}_\mathrm{th}\Gamma_\mathrm{M})\approx\SI{300}{\micro\s}$, which corresponds to the average time for one phonon to enter the high-$Q$ mode from the thermal bath.} Simultaneously, we achieve high scattered photon out-coupling efficiency of $\sim 70\%$ via properly engineered over-coupling. Similar systems have been used in experiments in the continuous-variable domain \cite{Moeller2017,Rossi2018, Nielsen2017}.

We characterize the optomechanical system by phase-modulating the drive light in a frequency range that covers the cavity resonance. This allows us to determine the cavity parameters, such as detuning $\Delta$ and optical linewidth $\kappa$. Furthermore, we use the same sweep to characterize the mechanical response based on the optomechanically-induced transparency (OMIT, see Ref. \cite{Weis2010}), which allows us to precisely determine the mechanical frequency $\Om$, light-enhanced coupling rate $g$ and optomechanical broadening $\Gopt$. These parameters, when combined with the knowledge of the outcoupling efficiency and the amount of transmitted optical power, allow us to estimate the mean intracavity photon number $\ncav$ and hence the single-photon coupling rate $g_0$.

\subsection{Filtering system}

The light emerging from the optomechanical cavity consists of the unscattered pump light (carrier), as well as the Stokes and anti-Stokes sidebands generated by mechanical motion. The sidebands generated by the mechanical mode of interest are not only close to the carrier frequency (merely \SI{1.48}{\MHz} away), but are also significantly weaker. The latter is due to the fact that the probability of Raman scattering for a mechanical resonator in the ground state is $4g_0^2/\kappa^2 < 10^{-8}$ \cite{Galland2014}. In addition, \orange{as will be further discussed in the Results section}, excited out-of-bandgap vibrational modes of the membrane are less than \SI{70}{\kHz} away from the mode of interest. 

The combinations of these factors places very stringent requirements on a filtering system that should be able to fully isolate the photons scattered by the mechanical mode of interest. Using a single Fabry-Pérot cavity for this task would not be practical: a cavity linewidth of \SI{300}{\Hz} would be needed to achieve the barely sufficient \SI{80}{\dB} of rejection at a detuning of \SI{1.5}{\MHz}. Apart from technical difficulties in implementation, such a filter would become increasingly inefficient as the optical mode becomes spectrally broader than \SI{300}{\Hz} due to dynamical backaction \cite{Aspelmeyer2014}, which is bound to happen during efficient optomechanical readout in our system. Finally, the $\sim \SI{0.5}{\ms}$ time delay introduced by a filter with \SI{300}{\Hz} bandwidth would exceed the estimated $\sim \SI{0.3}{\ms}$ decoherence time of our system, severely limiting our choices of experimental protocols.
 
The difficulties above can be avoided by using a cascade of several Fabry-Pérot cavities \cite{Riedinger2016,Cohen2015}. The rejection of a series of optical filters grows exponentially with the number of filters, being the product of the individual rejections, while the time delay scales only linearly with the number of cavities, being the sum of the individual delays. And importantly, the passband of the composite system remains sufficiently wide to accommodate broader signals. This approach allows us to use optical cavities with much more manageable linewidths.

Our filter system consists of four Fabry-Pérot cavities positioned in series, with each cavity having a linewidth of approximately \SI{30}{\kHz}. As described in more detail in Supplementary Material, the filter system is locked to a desired center frequency by sending an auxiliary locking beam to the system and sequentially locking the cavities. During measurement, the locking light is temporarily disabled to prevent saturation of the photon counter, and the photon counting of the filtered signal can take place. The intensity transmission of a single filter cavity, normalized to a peak value of 1, is given by the Lorentzian $L(\Omega)={1/[1+(2\Omega/\kf)^2]}$, where $\kf$ is filter cavity linewidth and $\Omega$ is detuning from the filter's resonance.
The complete filtering system consisting of four cavities is therefore expected to have a transmission of $L(\Omega)^4$.

The output of the optomechanical cavity is fiber-coupled to the input of the first filter as shown in Fig. \mref{fig:expset}{b}. At the end of the filter array we position a single-photon counting module (SPCM, avalanche photodiode COUNT-20C from Laser Components). The filter can then be tuned to observe either anti-Stokes ($\Df=\Om$) or Stokes ($\Df=-\Om$) emission. A mechanical shutter \cite{Zhang2015} is used to isolate the photon counter from the strong locking light present when the cavities are being actively locked.

\begin{figure}[t!]
	\includegraphics[width=1\linewidth]{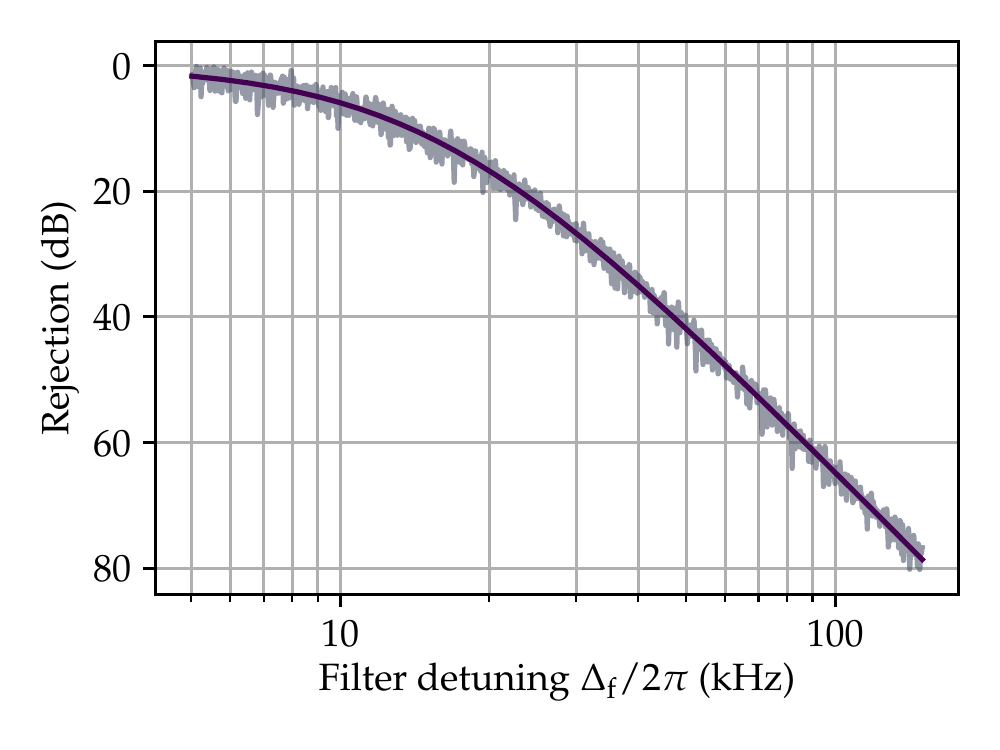}
	\caption{Rejection of the cascaded filtering system as a function of detuning, where the $L(\Omega)^4$ model (violet) closely follows the measured response (gray). For frequencies above \SI{200}{\kHz} the heterodyne signal becomes too weak to measure, but is expected to continue following the $L(\Omega)^4$ response.}
	\label{fig:filter_rejection}
	
\end{figure}

The characterization of the filter system is done at lower frequencies by applying phase-modulation sidebands on the incoming resonant light. By measuring the output from the filtering system with heterodyne detection, we recover both the resonant drive and the highly attenuated sidebands, which allows us to calculate the rejection of the whole filtering system with no need for additional calibration. As shown in Fig. \ref{fig:filter_rejection}, the rejection of the system is enough to strongly suppress ($>\SI{30}{\dB}$) the spurious mechanical modes which are only tens of \si{\kHz} away from the high-$Q$ mode. In addition, the strong optical drive detuned by the mechanical resonance frequency of $\approx \SI{1.5}{\MHz}$ is estimated to experience a much greater attenuation of $>\SI{155}{\dB}$, as given by the $L(\Omega)^4$ model. This makes the drive completely negligible compared to the scattering by the mechanical resonator.

For resonant light, the filter system has a transmission of $\sim 30\%$, currently limited mostly by losses at cavity mirrors and polarization optics. We believe that an overall transmission of 50\% should be possible by improving the cavity incoupling efficiencies and transmission between cavities, with further improvements requiring cavity mirrors with lower intrinsic losses.

\section{Results}

\subsection{Detection of filtered mechanical sidebands}
\begin{figure}[ht]
    \centering
    \includegraphics[width=1\linewidth]{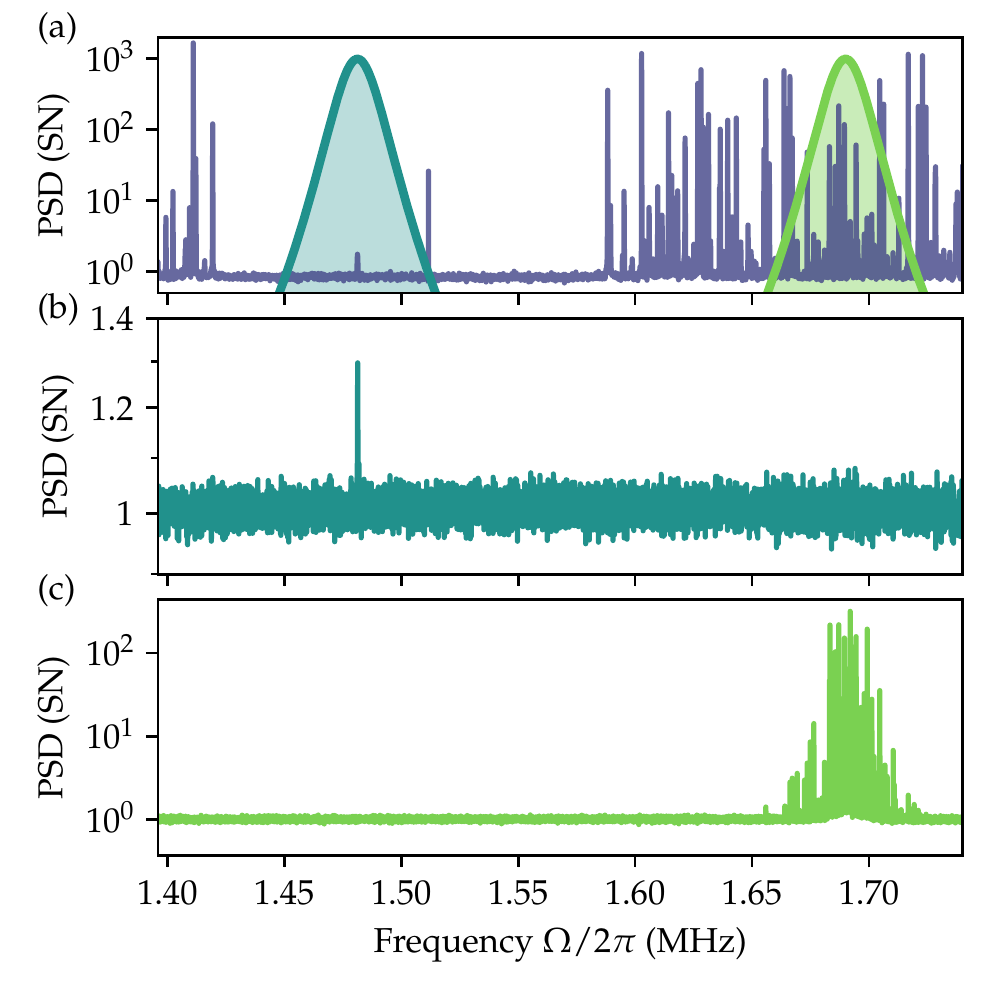}
    \caption{Filtering verified by heterodyne detection of filtered light. (a) Power spectral density (PSD) of light emitted directly from the cavity as registered by the direct-detection photodetector. PSD is calibrated in shot noise (SN) units. The spectrum shows the bandgap (between roughly $\SI{1.42}{\MHz}$ and $\SI{1.59}{\MHz}$) provided by the phononic crystal structure and the high-$Q$ mechanical defect mode at $\Om/2\pi=\SI{1.48}{\MHz}$. The overlaid shaded curves are transmission functions $L(\Omega-\Df)^4$ of the filtering system positioned at $\Df=\Om=2\pi\times\SI{1.48}{\MHz}$  and at $\Df=2\pi\times\SI{1.69}{\MHz}$. (b) and (c) show the PSD of filtered light, with the filter centered on the high-$Q$ mechanical defect mode frequency,
    (b) and on the dense out-of-bandgap part of the spectrum (c). Note the different scale between (b) and (c). 
    }
\label{fig:forest_filter}
\end{figure}

Unfiltered optomechanical spectra can be easily and efficiently measured using direct detection, i.e. using a single photodiode placed directly at the output of the optomechanical cavity. This is due to the fact that the cavity transduces membrane motion into light intensity fluctuations when the optical drive is correctly detuned \cite{Aspelmeyer2014}. As seen in Fig. \mref{fig:forest_filter}{a}, the spectrum consists of the main high-$Q$ mode surrounded by a phononic bandgap and dense regions of modes outside of the bandgap.

In order to confirm the performance of the filtering system, we apply it to the output of the OM cavity and detect the filtered light with a heterodyne measurement. The effect of filtering on the light spectrum is easily seen on Figs. \mref{fig:forest_filter}{b,c}, where we tune the filter to be resonant with distinct parts of the spectrum. When the filter is tuned to the main mechanical mode [Fig. \mref{fig:forest_filter}{b}], it efficiently isolates it from closely neighboring out-of-bandgap modes, which is a necessary condition for single-photon-based measurements. We can also select a part of the spectrum containing many out-of-bandgap modes, as in Fig. \mref{fig:forest_filter}{c}, which clearly reveals the envelope of the filter system's response.

\begin{figure}[ht]
    \centering
    \includegraphics[width=1\linewidth]{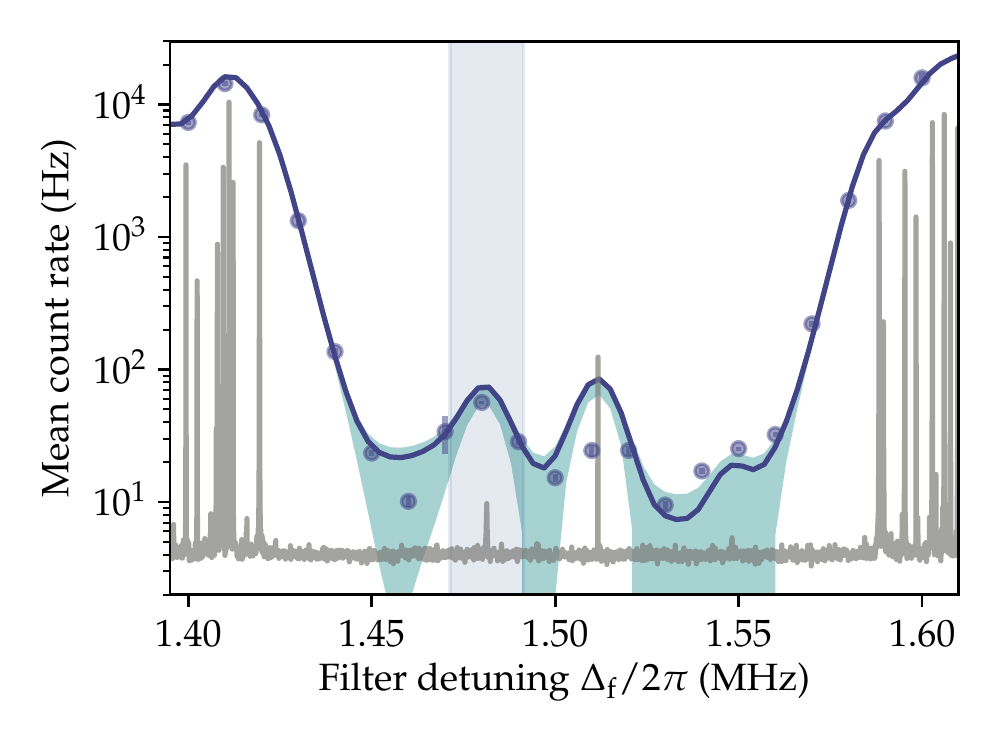}
    \caption{Photon counting of anti-Stokes sidebands, as a function of filter detuning, with count rates registered by the SPCM (blue dots), predicted count rates (solid blue line) with uncertainty (shaded teal area) and scaled directly-measured PSD (gray) for visual reference. The bandgap is effectively observed via photon counting. The shaded gray vertical strip shows the region surrounding the main mechanical mode. 
    }
    \label{fig:bandgap_scan}
\end{figure}

In a different measurement, we direct the output of the filtering system to the SPCM and lock the center frequency at different detunings across the bandgap (Fig. \ref{fig:bandgap_scan}). We observe greatly reduced photon scattering rates inside the bandgap, and large scattering when approaching low-$Q$ out-of-bandgap modes that are strongly coupled to the thermal bath. As a consistency check, we estimate the expected count rates by convolving the $L(\Omega)^4$ response of the filters with a directly measured spectrum, where shot noise has been subtracted. The predicted and measured rates are in good agreement, with visible uncertainty only inside of the bandgap, where the scattering is low and shot noise level estimation errors can lead to increased uncertainties in the predicted count rate, as shown in Fig. \ref{fig:bandgap_scan}. \orange{The residual discrepancy at 1.51 MHz is due to one of the higher-order defect modes that is very weakly damped by light and prone to mechanical excitation due to e.g. unstable liquid helium flow through the cryostat. This mode does not affect the measurements of the main mode, as shown on Fig. \mref{fig:forest_filter}{b}, as it is being suppressed by the filter systems by \SI{30}{\dB} (see Fig. \ref{fig:filter_rejection})}. Importantly, when the filters are tuned to the main mechanical frequency, the relative photon flux contribution of out-of-bandgap modes is expected to be less than 1\% compared to the photon flux due to the main mechanical mode.

\subsection{Raman-ratio phonon thermometry}
An asymmetry between Stokes and anti-Stokes sidebands is a direct signature of near-zero occupation of the mechanical mode responsible for these sidebands. In particular, the ratio of powers in the two sidebands can be directly used to infer the residual phonon occupation or, equivalently, the mode temperature. For mechanical resonators in the \si{\MHz} frequency range, the most commonly employed method for measuring sideband powers has been heterodyne detection \cite{Purdy2015,Peterson2016,Underwood2015,Chowdhury2019}. In particular, it is fully sufficient to perform heterodyning of photons scattered from red-detuned cooling light. In this case, as cooling increases, the two sidebands move from being asymmetric due to the optomechanical cavity response, to being equally strong, indicating the balance of Stokes and anti-Stokes scattering in the quantum back-action dominated regime.

In the \si{\GHz} mechanical frequency range, a more direct method of measuring sideband power based on photon counting (and thus effectively phonon counting) \cite{Cohen2015,Meenehan2015} has been demonstrated, where Stokes and anti-Stokes photons are filtered and subsequently detected by a single-photon detector.

This method is not affected by the local oscillator noise \cite{Weinstein2014}, although it may suffer from dark counts of the photon counting detectors \cite{Cohen2015}. 
The calibration-free nature of Raman-ratio thermometry, in both resonant and red-detuned cases, is one of its advantages as compared to the more commonly employed  technique based on spectral calibration using external phase modulation \cite{Gorodetsky2010}. In particular, one does not need to pre-calibrate the optomechanical single-photon coupling rate strength $g_0$.  In the context of our work, phonon counting thermometry demonstrates the feasibility of efficient counting of single-phonon excitations, a fundamentally non-Gaussian operation. 

Here we apply the phonon counting thermometry technique to our $\Om/2\pi = \SI{1.48}{\MHz}$-frequency mechanical mode. A single beam detuned from the optomechanical cavity resonance by approximately optimal detuning $\Delta/2\pi= \SI{-1.85}{\MHz}$, is used to both cool the membrane motion by dynamical back-action and simultaneously probe the system as it reaches the quantum back-action dominated regime.

\begin{figure}[ht!]
    \centering
    \includegraphics[width=1\linewidth]{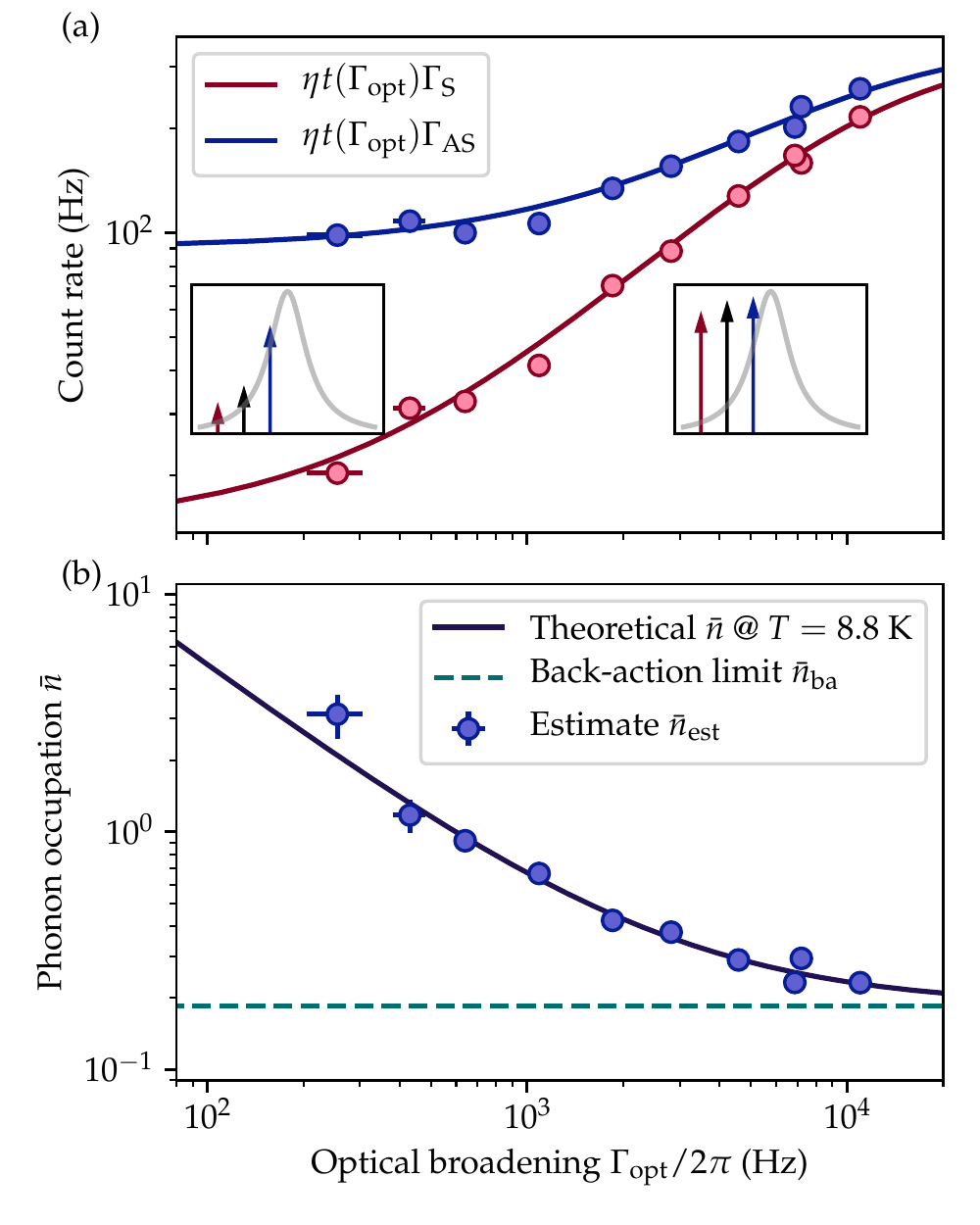}
    \caption{Ground-state cooling measured by photon counting. (a) Measured Stokes (red) and anti-Stokes (blue) scattering rates as a function of optical broadening $\Gopt$ of the mechanical mode and theoretical prediction with calibrated efficiency, as given in the legend. At lower driving powers, corresponding to smaller broadenings, the rates are set by the Lorentzian cavity response (left inset). At higher driving powers we observe that asymmetry is reduced due to the mechanical oscillator approaching the ground state (right inset), with scattering dominated by the quantum back-action. (b) Inferred thermal occupation $\nest$ of the mechanical mode, along with the theoretical prediction (for bulk thermalization temperature of $T = \SI{8.8\pm0.5}{\K}$, which is the only free parameter in the fit) and the back-action limit $\nba$ (dashed horizontal line). Error bars are inferred from statistical uncertainties from photon counting and fitting of other parameters used in Eq.~(\ref{eq:nest}).}
    \label{fig:asymmetry}
\end{figure}
The transition rates for the mechanical system can be calculated following Refs. \cite{Wilson2007,PhysRevLett.99.093902,Aspelmeyer2014} as:
\begin{equation}
     A_\pm = g_0^2 \ncav\frac{ \kappa }{(\Delta \mp \Om)^2+\frac{\kappa ^2}{4}},
\end{equation}
where $+$ ($-$) denotes upward (downward) transitions in the quantum harmonic oscillator ladder. The expected Stokes and anti-Stokes rates are then given by:
\begin{equation}
    \GAS = \bar{n} A_-,\quad    \GS = (\bar{n} +1) A_+,
    \label{eq:rawrates}
\end{equation}
with the dynamical optical broadening given by ${\Gopt = A_- - A_+}$. Remarkably, in the ground state ($\bar{n}=0$) the rates become highly asymmetric, regardless of $A_\pm$. The ratio between Stokes and anti-Stokes count rates is unaffected by the overall system efficiency, thus we can estimate the residual phonon occupancy from these rates as:
\begin{equation}
\begin{aligned}
    \nest&=\frac{R A_+}{A_- - R A_+}\\
    & =\frac{R ((\Delta+\Om)^2+\kappa^2/4)}{((\Delta-\Om)^2+\kappa^2/4) - R ((\Delta+\Om)^2+\kappa^2/4)},
\end{aligned}
\label{eq:nest}
\end{equation}
with $R=\GAS/\GS$. Notably, both $g_0$ and the cavity photon number $\ncav$ cancel out in the estimator. Other parameters of the cavity are found via OMIT measurements, as described in the Sec.~2.\ref{sec:omset}. The theoretical prediction for the expected final phonon occupancy can be calculated as:
\begin{equation}
    \bar{n} = \frac{A_+ + \nth \Gm}{\Gopt+\Gm}.
    \label{eq:ntheo}
\end{equation}
In our case of $\Gopt\gg\Gm$, two regimes can be distinguished. In the thermally-dominated regime, corresponding to $\Cq \ll 1$, the ratio between the rates is determined only by the cavity response, leading to ${R\rightarrow ((\Delta - \Om)^2+\kappa ^2/4)/((\Delta + \Om)^2+\kappa ^2/4)}$ and ${\GAS=\nth\Gm (1-1/R)^{-1}}$. In the sideband-resolved regime, $R\rightarrow\infty$ and the anti-Stokes rate becomes equal to the phonon flux into the mechanical resonator coming from the thermal bath. In the other extreme (i.e. $\Cq\gg1$) the two rates equalize and we have $\GS\approx\GAS\rightarrow g_0^2 \ncav \kappa/(4 |\Delta| \Om)\propto \Cq$, which shows that the scattering rates of two processes become equal and are dominated by quantum back-action.

We now proceed to demonstrate this behavior in our experimental setting.
 Figure \mref{fig:asymmetry}{a} demonstrates the measured Stokes and anti-Stokes rates both growing with the intracavity photon number, quantified in terms of induced optical broadening $\Gopt$ measured using OMIT. 
 For the lowest broadening of $\Gopt/2\pi=\SI{255}{\hertz}$ we observe sideband scattering rates of $\SI{20}{\hertz}$ for Stokes and $\SI{100}{\hertz}$ for anti-Stokes, corresponding the the ratio $R=5$. As we increase the broadening to $\Gopt/2\pi=\SI{11}{\kilo\hertz}$, the detected rates arrive at $\SI{215}{\hertz}$ for Stokes and $\SI{260}{\hertz}$ for anti-Stokes corresponding to a ratio of $R=1.2$.
 Increased optical broadening also leads to reduced transmission through the filter setup, as compared with raw rates given by Eq.~(\ref{eq:rawrates}), which is due to the optically-broadened scattered light getting slightly "clipped" by the filtering system response. We model this loss by integrating a normalized Lorentzian spectrum with a width $\Gopt$ centered around $\Om$ with $L(\Omega-\Om)^4$:
\begin{equation}
\begin{aligned}
t(\Gopt) &= \int L(\Omega-\Om)^4 \frac{2}{\pi\Gopt}\frac{\Gopt^2/4}{(\Omega-\Omega_m)^2+\Gopt^2/4} \mathrm{d}\Omega  \\&=\frac{\kf  \left(5 \Gopt ^3+20 \Gopt^2 \kf +29 \Gopt  \kf ^2+16 \kf ^3\right)}{16 (\Gopt+\kf )^4}.
\end{aligned}
\end{equation}
This reduction is the same for Stokes and anti-Stokes sidebands, and thus it does not affect the ratio $R$. We find that the optical spring effect shifting the effective mechanical resonance frequency is below \SI{3}{\kHz} has a negligible effect on count rates for a fixed detuning of the filter system. In all cases we subtract the independently measured dark count rate of \SI{15.5(5)}{\Hz}. The overall detection efficiency of the entire system is estimated to be $\eta\approx2.5\%$, consisting of optomechanical cavity outcoupling ($75\%$), fiber transmission/coupling ($60\%$), filtering system ($30\%$ for cavities and $50\%$ for incoupling/outcoupling) and SPCM efficiency ($35\%$). We note that the room for improvement of these efficiencies lies mostly in optics of the filtering system and SPCM efficiency.

Finally, we calculate the ratio $R$ and estimate the mean phonon occupancy as given by Eq.~(\ref{eq:nest}) and plotted in Fig. \mref{fig:asymmetry}{b}. The estimated phonon occupation $\nest$ is accurately described following a fit of Eq.~(\ref{eq:ntheo}). The mechanical occupation finally reaches a value of ${\nest=0.23\pm0.02}$ at ${\Gopt/2\pi = \SI{11.0}{\kHz}}$ corresponding to ${\Cq\approx22}$, as estimated from calibrated parameters.  The only free parameter of the theory is the the temperature of the phononic bath, determined to be $T=\SI{8.8\pm0.5}{\K}$, which is consistent with previous works involving similar mechanical systems \cite{Rossi2018}. The minimum occupation achievable with sideband cooling, often referred to as the back-action limit, lies at ${\nba=(A_-/A_+-1)^{-1}\approx0.185}$ for our case. We thus observe a strong suppression of the classical sideband asymmetry due to the mechanical oscillator motion being primarily driven by the radiation-pressure shot noise.

\subsection{Phonon correlation interferometry}
Lastly, we concentrate on statistical properties of light emitted from the high-$Q$ mechanical mode. We set $\Gopt/2\pi=\SI{2.1}{\kHz}$ and park the filter at the anti-Stokes sideband.  We collect a total of $\num{18d3}$ counts at a count rate of \SI{90}{\Hz}, and look at coincidences between counts as a function of the delay time~$\tau$.  Since we only use a single detector, we reject the events for which $|\tau| < \SI{500}{\ns}$, in order to avoid effects of dead time of the SPCM and afterpulsing. \orange{This time is still much shorter than any dynamics present in the system, and thus we can extrapolate our results on the coincidence rate to the zero-delay value.} We analyze the coincidences in terms of the second-order Glauber correlation function ${\gtwo(\tau)=\langle \hat{a}^\dagger(0)\hat{a}^\dagger(\tau) \hat{a}(0)\hat{a}(\tau)\rangle/\langle \hat{a}^\dagger(0) \hat{a}(0) \rangle \langle \hat{a}^\dagger(\tau) \hat{a}(\tau) \rangle}$. 

\begin{figure}[t]
    \centering
    \includegraphics[width=1\linewidth]{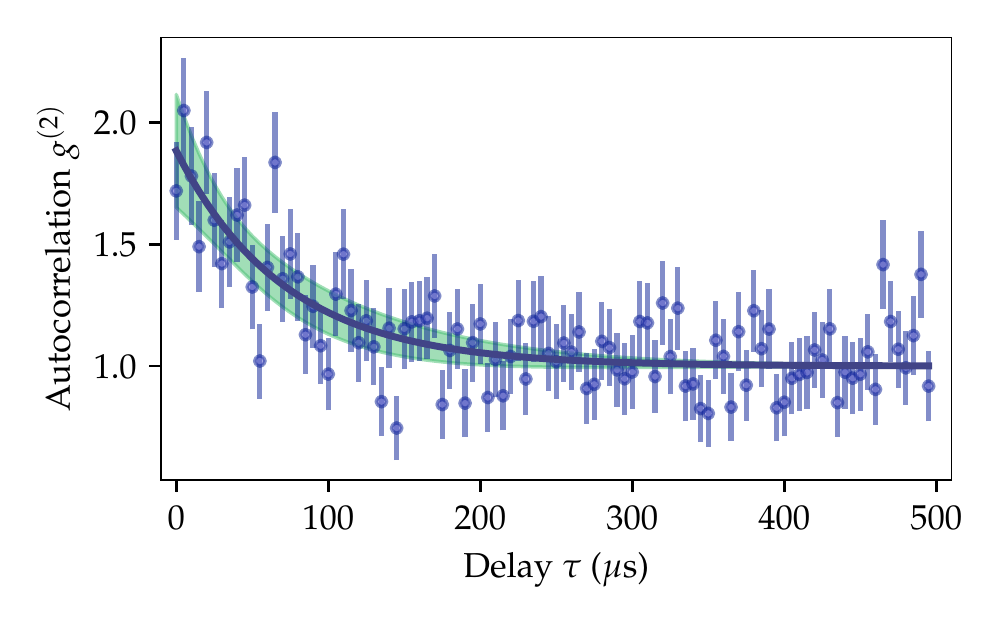}
    \caption{Second-order autocorrelation measurements of the spectrally-filtered anti-Stokes photons from the high-Q mechanical mode. We estimate ${\gtwo(0)=1.88\pm0.08}$ and optical coherence time $\tau_\text{C}=\SI{143(18)}{\us}$. Error bars in the plot are inferred from statistical uncertainties of Poissonian counts, while shading for the fitted curve corresponds to three s.d. confidence bounds.}
    \label{fig:g2auto}
\end{figure}

Light scattered by a single mechanical mode in thermal equilibrium has thermal statistics, as described by the following second-order correlation function:
\begin{equation}\label{eq:g2}
    \gtwo(\tau) = 1+A \exp(-2|\tau|/\tau_\text{C})=1+A \exp(-\Gopt|\tau|),
\end{equation}
where $\tau_\text{C}=2/\Gopt$ is the coherence time of light, with $\gtwo(0)=2$ \orange{for $A=1$}.
For a multimode thermal state, one would expect a multi-exponential or oscillatory decay. Our measurement, shown in Fig.~\ref{fig:g2auto}, shows a single-exponential decay with a decay time matching the optically-broadened linewidth and exhibits $\gtwo(0)=1.88\pm0.08$ (obtained from a fit of Eq.~(\ref{eq:g2}) \orange{with $A$ and $\tau_\mathrm{C}$ as free parameters}), which is close to the theoretical value of 2, indicating the high purity and single-mode behavior of the measured thermal state of light. The optical coherence time of $\tau_\text{C}=\SI{143(18)}{\us}$ (corresponding to an optical linewidth of \SI{2.2(3)}{\kHz}), closely matches the optical broadening of the mechanical oscillator, independently measured by OMIT to be $\Gopt/2\pi = \SI{2.1}{\kHz}$. This feature confirms our system's potential for producing non-Gaussian quantum states of light and of motional degrees of freedom. We attribute the residual discrepancy to the dark counts that exhibit Poissonian counting statistics.

\section{Conclusions and Outlook}
We have demonstrated a versatile optomechanical system where an ultracoherent high-$Q$ mechanical resonator is subject to both discrete and continuous variable measurements, paving the road towards generation of a wide range of non-classical states of motion. 
We have directly demonstrated selection of photons scattered from a single mechanical mode by heterodyne spectroscopy, as well as by second-order single-photon intensity interferometry. The mode has been optically cooled to a final phonon occupation of $0.23\pm0.02$, which has been measured using Raman-ratio thermometry via photon counting. Our work marks the first application of phonon counting techniques to low-frequency mechanical resonators, paving the way towards generation of non-Gaussian mechanical states \cite{Galland2014,Khalili2010} and studying related decoherence processes \cite{Pepper2012,Weaver2018,Sekatski2014}, \orange{for which the relatively larger mass of our system, as compared to GHz-frequency resonators \cite{Ren2019, Hong2017}, is a} \blue{clear asset.}

\orange{The path towards generating non-Gaussian macroscopic quantum states} \blue{in our system} \orange{presents \blue{additional} technical challenges. The mean phonon occupation of $0.23$ demonstrated here is at best borderline for demonstrating non-classical features \cite{Galland2014}. The main step will be employing a narrower optomechanical cavity, to simultaneously allow better sideband cooling to at least $0.1$ phonons in the fully sideband-resolved regime ($\kappa/2\pi \sim \SI{300}{\kilo\hertz}$ for which $\nba\sim0.003$) and high degree of selectivity of Stokes or anti-Stokes processes. In such a regime, the technical noise of the laser is expected to start limiting the occupation \cite{Safavi-Naeini2013}. Furthermore, performing Raman-ratio thermometry will require photon detectors with very low ($<1/\mathrm{s}$) dark count rates.}

The ultra-narrowband filtering technique we developed can become useful in many different optomechanical systems, ranging from other low-frequency devices such as trampoline resonators \cite{Norte2016}, to macroscopic levitated particles \cite{Tebbenjohanns2019}, as well as atomic ensembles \cite{Lvovsky2009,Hammerer2010} and ionic or defect emitters in solid state \cite{Awschalom2018}.
In the case of atomic ensembles, optical cavities are routinely used to distill weak quantum light \cite{Palittapongarnpim2012}, but ultra-narrowband filters, such as demonstrated here, would be required to employ photon counting techniques for quantum memories operating in the spin-exchange relaxation free (SERF) regime \cite{Katz2013,Katz2018} or based on motional averaging \cite{Borregaard2016,Zugenmaier2018}, for example. Narrowband filtering can be beneficial for solid-state emitters as well, allowing better understanding of their optical properties \cite{Perrot2013,Bartholomew2017}, as well as enabling spectrally-based selection of single emitters from an ensemble \cite{Casabone2018}.
 
Narrowband photonic states demonstrated here can be directly interfaced with material systems of long coherence times, facilitating long-distance quantum communication. Ultimately, they can also be used in hybrid quantum networks \cite{Moeller2017,Thomas2020} to generate entangled states via heralded photon counting.
 
\medskip
\noindent\textbf{Funding information.}
ERC Advanced Grant QUANTUM-N and Villum Investigator Grant QMAC.  

\noindent\textbf{Acknowledgments.}
We acknowledge contributions to soft-clamped membrane design by A. Schliesser, discussions with J. Appel, J.~H. Müller, M. Zugenmaier, K.~B. Dideriksen, B. Albrecht, contributions to laser stabilization and electronic design by T. Zwettler, technical assistance by D. Wistisen,  and early-stage development of the experiment by A. Barg. M.~P. was partially supported by the Foundation for Polish Science (FNP).

\noindent\textbf{Disclosures.} The authors declare no conflicts of interest.

\bigskip \noindent See Supplementary Material for supporting content.

\bibliography{bibliography}

\clearpage
\onecolumngrid
\section*{Supplementary Material}
\twocolumngrid
\section{Implementation of the filtering system}

\subsection{Construction and material}
\begin{figure}[ht!]
	\includegraphics[width=1\linewidth]{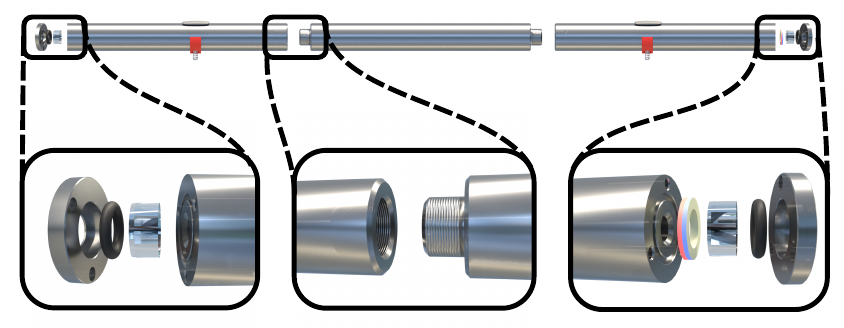}
	\caption{Design of a filtering cavity. The resonator's main body (top) is composed of 3 parts joined by threads (central inset). One of the mirrors is pressed directly against the main body (left inset), while the other is positioned on a piezo-transducer to allow for cavity locking (right inset).}
	\label{fig:filter_mechanical}
\end{figure}

We have chosen Invar 36 (1.3912) as the material for the construction of the spacers for our filtering resonators. Invar provides a low coefficient of thermal expansion of $<\SI{1.5}{ppm/\K}$ \cite{invar} at room temperature, while being significantly easier to machine, more robust, available and economic than higher-grade low-expansion materials such as ULE glass \cite{Alnis2008,Argence2012}. In order to achieve a length of \SI{60}{\cm}, we have constructed each cavity out of three pieces that are joined by two threaded connections as shown in Fig. \ref{fig:filter_mechanical}.

\subsection{Vibration isolation}
\begin{figure}[ht!]
	\includegraphics[width=1\linewidth]{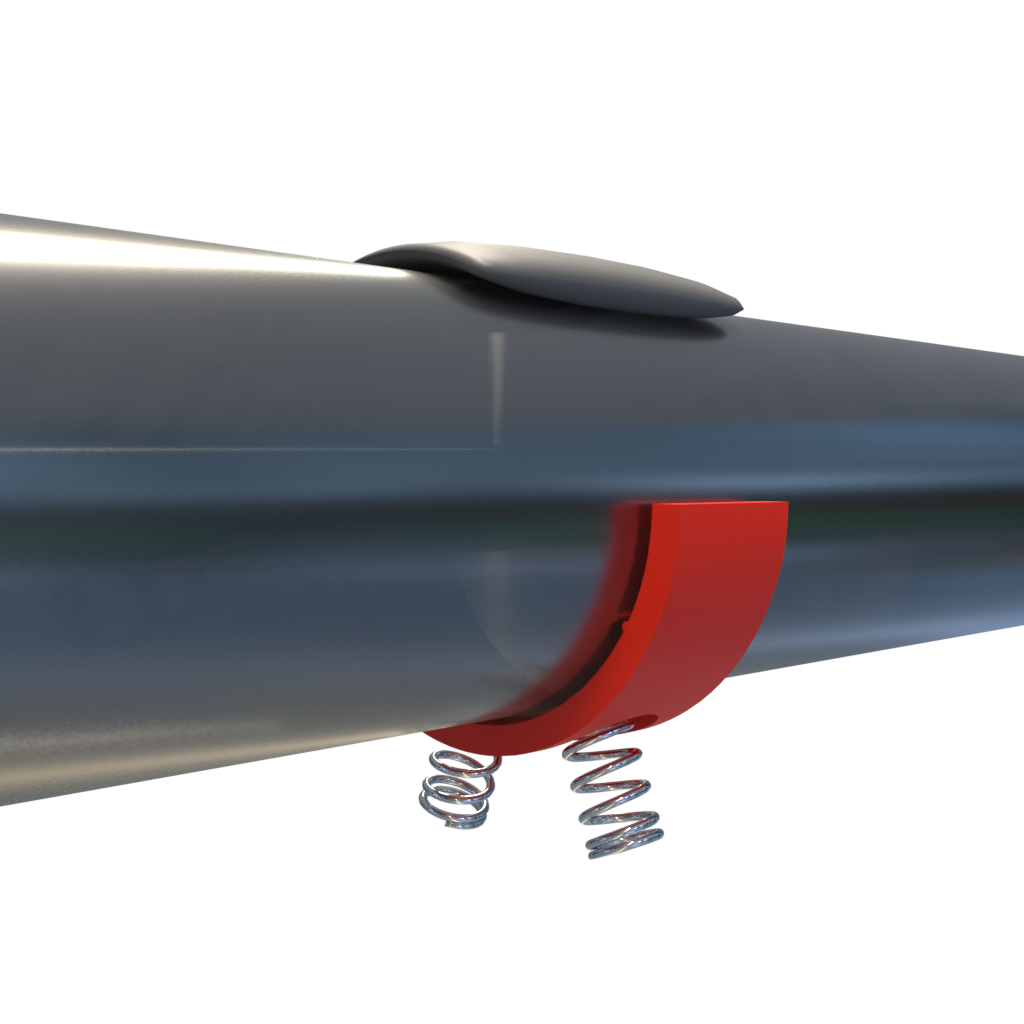}
	\caption{Damped spring cavity support. The two soft springs on the bottom give the cavity-spring system a resonance frequency of \SI{10}{\Hz} and block external high-frequency vibrations. The cellulose damper (top of the cavity) dampens the remaining \SI{10}{\Hz} cavity-spring vibrations. This damper is tightly inserted between the top of the filter cavity and the inner surface of the vacuum tube.}
	\label{fig:damper}
\end{figure}

The cavities are placed inside four long cylindrical vacuum tubes (KF40) that are part of the same vacuum system. Given the substantial length of each filtering resonator, they are susceptible to vibrations at their first and second bending mode frequencies (\SI{260}{\Hz} and \SI{720}{\Hz} respectively).  To isolate the system from external perturbations, we employ damped spring supports. There, each cavity is supported from below at its Airy points by two mounts of two soft springs each (\SI{5}{\N/\mm}), as shown in Fig. \ref{fig:damper}. This method gives a cavity-spring oscillation frequency of approximately \SI{10}{\Hz} and translates into a rejection of more than \SI{52}{\dB} of power of external vibrations at the frequency of the first flexural mode of the spacer. To dampen the remaining \SI{10}{\Hz} oscillations, we employ a cellulose (cotton) layer on top of each cavity, has proven to have minimal outgassing in our vacuum conditions of \SI{1d-4}{\hecto\Pa}. The resulting stability and vibration immunity is excellent, allowing normal alignment work to be performed on any given cavity while locking other cavities and using light transmitted through them.

\subsection{Dither-locking}
\begin{figure}[ht]
	\includegraphics[width=1\linewidth]{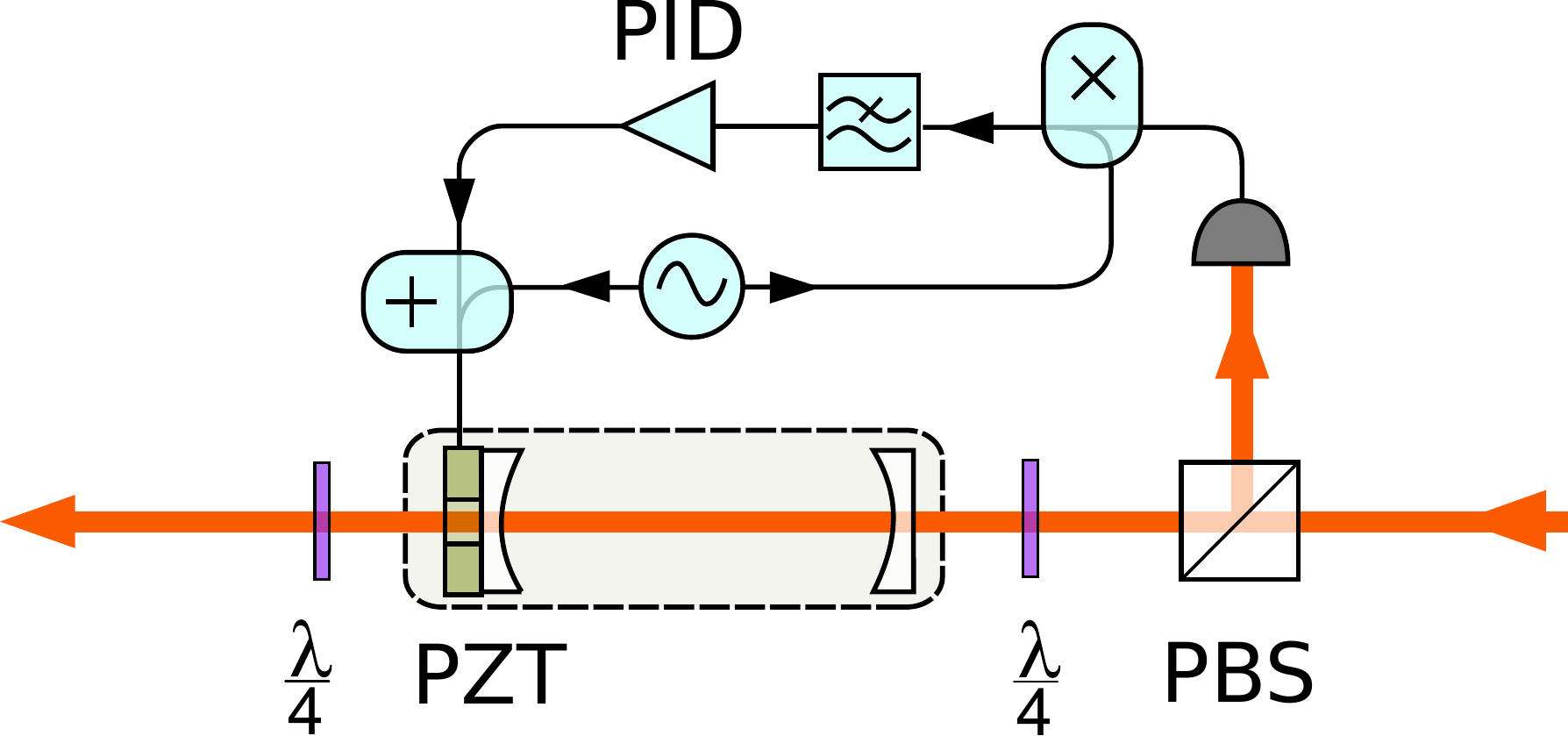}
	\caption{Dither-lock scheme for filter cavities. The light reflected from the cavity is redirected towards the photodetector. The length of the cavity is modulated by the piezoelectric transducer (PZT). The resulting reflection signal is digitally demodulated and used to lock the resonator on resonance.}
	\label{fig:dither_diagram}
\end{figure}

In order to lock all four cavities on resonance, we use the well-known technique of dither-locking, functionally equivalent to low-frequency Pound-Drever-Hall locking \cite{pdh}. Here, a periodic $\sim \SI{1}{\kHz}$ modulation is applied to each filter's piezoelectric transducer, therefore modulating each cavity's resonance frequency.  The reflected light intensity is detected and demodulated, as shown in Fig.~\ref{fig:dither_diagram}. This produces an error signal that is directly related to the derivative of each cavity's reflection function. By using slightly different modulation frequencies for different cavities, interference of signals between the different dither drives is avoided. This technique, when compared to the conventional Pound-Drever-Hall scheme, has the advantage of not requiring an electro-optical modulator before each cavity, which would otherwise introduce significant losses and complexity into the optical setup. Low-level locking functionality, including modulation, demodulation, and feedback, is implemented digitally in ARM-based microcontroller boards (Arduino Due) attached to each filter cavity, while high-level coordination of the four microcontroller boards is performed by a Python application running on computer. 

The full locking procedure consists of the following steps. In the beginning, the first cavity is scanned using its piezo transducer in order to find several of its TEM00 fundamental optical resonances (3 to 4 free spectral ranges are scanned).  Locking light is generated from a common laser with a pair of frequency-tuned acousto-optic modulators (AOMs). Once the optical resonances have been identified, the first cavity is locked on the side of one of those resonances, i.e. with a detuning of $\Delta = \kf/2$. Once lock is confirmed to be stable, dithering is enabled and the new dither-derived error signal is used to lock the cavity on its resonance ($\Delta = 0$). With light being transmitted through the first cavity, the next cavity is now locked via the same procedure, and so on. While lock light is being used, one mechanical shutter \cite{Zhang2015} blocks this light from going to the fragile SPCM, and a second shutter prevents any spurious back-reflections of lock light from propagating into the sensitive optomechanical system. 

When performing photon-counting measurements, lock light is disabled by turning off the AOM drives that generate it, and the locks of the filter cavities are switched into the dead-reckoning (,,frozen lock'') mode. The optical paths to the SPCM and the optomechanical setup are subsequently opened to allow the signal to propagate through the filters.

After several seconds of passive stability and photon collection, we close the mechanical shutter to block the input to the SPCM and re-enable the locking light. The dither-locking is then re-engaged and the four cavities, still being in the vicinity to their resonances, are promptly brought back to resonance within less than \SI{500}{\ms}, as shown in Fig. \ref{fig:passive_stability}.

\subsection{Measurement of passive stability}
To be useful in pulsed protocols involving photon counting, the filtering setup must be stable enough to operate without any reference light input ("frozen lock" regime) while photons are being counted. In this regime, active piezo feedback is paused and the reference lock light is disabled. The cavities' length then evolves freely, and it is crucial that all cavities remain sufficiently close to their resonances during the relevant timescale. 

\begin{figure}[ht!]
    \centering
	\includegraphics[width=1.0\columnwidth]{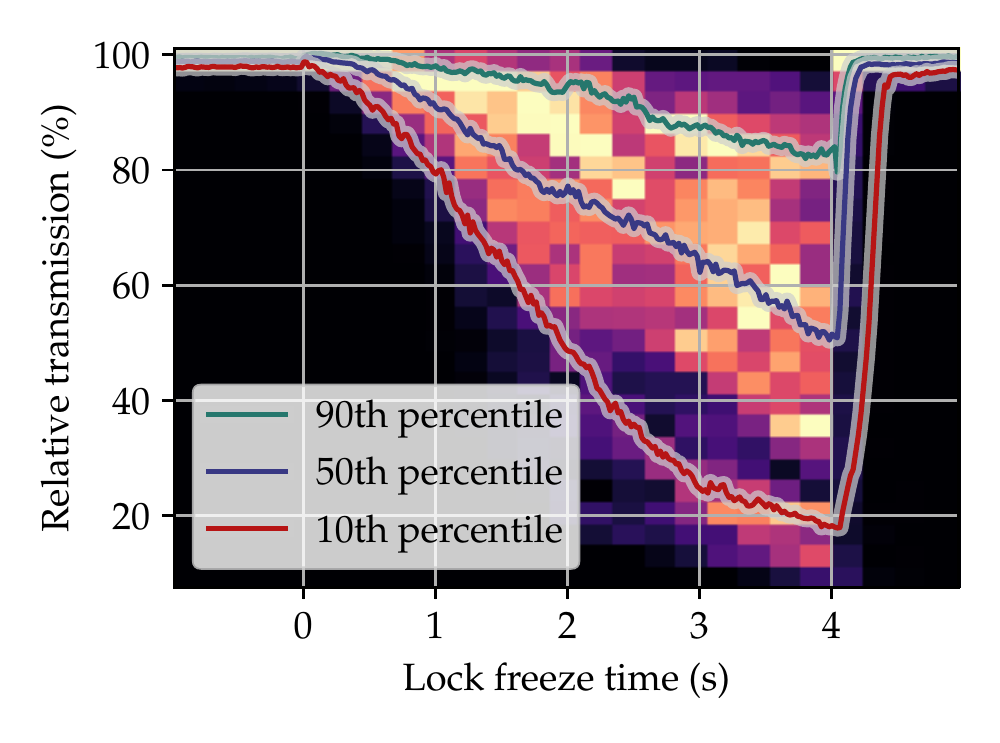}
	\caption{Transmission of the four-cavity filtering system over \SI{4}{\s} after freezing the locking feedback loop. Feedback is active for times before \SI{0}{\s} and after \SI{4}{\s}, showing the high passive stability of the system during dead-reckoning. The background color plot corresponds to the probability density of relative transmission.}
	\label{fig:passive_stability}
\end{figure}

On average, the relative transmission of the entire filter system decays to 50\% in more than 4 seconds, as shown in Fig.~\ref{fig:passive_stability}, and in 90\% of the cases, the relative transmission of the full filter system stays above 80\% for approximately 1 second. Importantly, 1 second is more than $3000$ times longer than the next slowest experimental timescale, namely the \SI{0.3}{\ms} decoherence time of our mechanical resonator, given by $T_1 = \hbar Q / (\kB T)$ \cite{Aspelmeyer2014}, where we assume a bath temperature of \SI{9}{\K} and a $Q$-factor of \num{380d6}.

\orange{During data acquisition, a lock freezing time of \SI{1.5}{\s} was used. With this setting, the relative efficiency of the filters is $92 \pm 6 \%$ (averaged over the freezing interval). Notably, a typical experimental run is comprised of many lock-freeze cycles, which permits us to average out technical fluctuations (e.g. filter system transmission) and Poissonian count rate fluctuations. Statistical properties of the ensemble of these cycles then allow us to accurately estimate the uncertainty on photon rates, as well as dark count rate.}

\end{document}